\documentclass[twocolumn,noshowpacs,amsmath,amstex,amssymb,mathfonts,prl]{revtex4}
\usepackage{amsthm,amsfonts,graphicx,verbatim}
\usepackage{amsmath}
\usepackage{amssymb}
\usepackage{amsthm}
\usepackage{amsfonts}
\usepackage{listings}
\usepackage{enumerate}
\usepackage{latexsym}
\usepackage{psfrag}
\usepackage{bm}
\usepackage{graphicx}
\usepackage{subfigure}
\usepackage{float}

\usepackage{color}

\newcommand{\si}{Supplementary Information}

\newcommand{\be}{\begin{equation}}
\newcommand{\ee}{\end{equation}}
\newcommand{\bea}{\begin{eqnarray}}
\newcommand{\eea}{\end{eqnarray}}

\newcommand{\la}{\langle}
\newcommand{\ra}{\rangle}

\renewcommand{\epsilon}{\varepsilon}

\begin{document}

\title{Direct Measurement of the Zak phase in Topological Bloch Bands}

\author{Marcos Atala,$^{1,\dagger}$ Monika Aidelsburger,$^{1,\dagger}$ Julio T. Barreiro,$^{1,2}$\\ Dmitry Abanin,$^{3}$ Takuya Kitagawa,$^{3}$ Eugene Demler,$^{3}$ Immanuel Bloch$^{1,2}$}
\affiliation{%
$^{1}$Fakult\"at f\"ur Physik, Ludwig-Maximilians-Universit\"at, Schellingstr. 4, 80799 Munich, Germany\\
$^{2}$Max Planck Institute of Quantum Optics, Hans-Kopfermann Str. 1, 85748 Garching, Germany\\
$^{3}$Department of Physics, Harvard University, 17 Oxford Str., Cambridge, MA 02138, USA\\
$^\dagger$These two authors contributed equally to this work.}

\begin{abstract}
Geometric phases that characterize the topological properties of Bloch bands play a fundamental role in the modern band theory of solids. Here we report on the direct measurement of the geometric phase acquired by cold atoms moving in one-dimensional optical lattices. Using a combination of Bloch oscillations and Ramsey interferometry, we extract the Zak phase -- the Berry phase acquired during an adiabatic motion of a particle across the Brillouin zone -- which can be viewed as an invariant characterizing the topological properties of the band. For a dimerized optical lattice, which models polyacetylene, we measure a difference of the Zak phase equal to $\delta\varphi_{\text{Zak}}=0.97(2) \pi$ for the two possible polyacetylene phases with different dimerization. This indicates that the two dimerized phases belong to different topological classes, such that for a filled band, domain walls have fractional quantum numbers. Our work establishes a new general approach for probing the topological structure of Bloch bands in optical lattices. 
\end{abstract}
\maketitle

The non-trivial topological structure of Bloch bands in solids gives rise to fundamental physical phenomena, including fermion number fractionalization~\cite{Jackiw76,Wilczek,SSH}, the quantum Hall effect~\cite{TKNN,NiuReview}, as well as topologically protected surface states in topological insulators~\cite{KaneHasan,Qi:2011}. The topological character of a Bloch band is defined by certain invariants, which can be expressed in terms of the Berry's phase~\cite{Berry} acquired by a particle during adiabatic motion through the band~\cite{NiuReview,Qi:2011}. The most well-known example is the two-dimensional topological invariant, the first Chern number, which is related to the Berry's phase for a contour enclosing the Brillouin zone and determines the quantized value of the Hall conductivity of a filled two-dimensional band~\cite{TKNN,NiuReview}. For one-dimensional systems, topological invariants of Bloch bands have been discussed theoretically~\cite{Zak89,NiuReview,Qi:2011}, however never been measured in any experiment.
\\

Here we present direct measurements of Berry's phase and topological invariants of one-dimensional periodic potentials using systems of ultra-cold atoms in optical lattices. Topological properties of one dimensional solids are characterized by the so-called Zak phase -- the Berry's phase picked up by a particle moving across the Brillouin zone~\cite{Zak89}. For a given Bloch wave $\psi_{k}(x)$ with quasimomentum $k$, the Zak phase can be conveniently expressed through the cell-periodic Bloch function $u_k(x)=e^{-ikx}\psi_k(x)$: 
\begin{align}
\varphi_\text{Zak} &= i \int_{-G/2}^{G/2} \langle u_k |\partial_k | u_k \rangle \, dk, 
\label{eq:Zak}
\end{align} 
where $G=2\pi/d$ is the reciprocal lattice vector and $d$ is the lattice period~\cite{Zak89}. Non-trivial Zak phases underlie the existence of protected edge states~\cite{Ryu02,Montambaux11}, fermion number fractionalization~\cite{Jackiw76,Wilczek,SSH}, and irrationally charged domain walls~\cite{NiemiSemenoff,RiceMele} between topologically distinct one-dimensional solids. These phenomena, initially discussed in the context of quantum field theory~\cite{Jackiw76,NiemiSemenoff,Wilczek}, later on found condensed matter realizations in polyacetylene~\cite{SSH}, described by the celebrated Su-Schrieffer-Heeger (SSH) model, and linearly conjugated diatomic polymers~\cite{RiceMele}. \\

\begin{figure}[h]
\begin{center}
\includegraphics[width=0.9\linewidth]{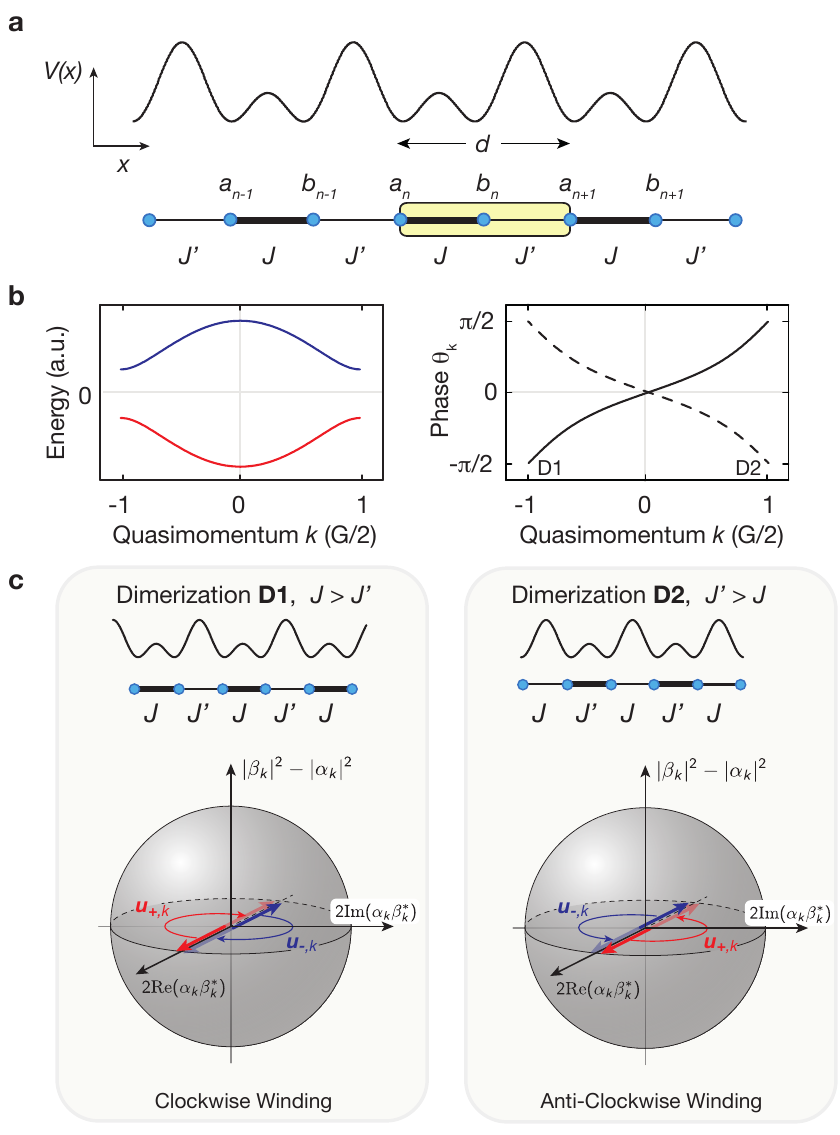}
\end{center}
\caption{Energy bands and topology of dimerized lattice model. {\bf (a)} Schematic illustration of optical superlattice potential used in the experiment to realize the Su-Schrieffer-Heeger model (yellow box denotes the unit cell of size $d=\lambda_s$). {\bf (b)} Exemplary curves for the lower and upper energy bands (red and blue lines) and phase $\theta_k$ for dimerization D1 and D2 (solid and dashed line) as a function of quasimomentum $k$. {\bf (c)} Pseudo-spin representation of the eigenstates $\mathbf u_{\mp,k}$~of the upper and lower energy bands for the two dimerization configurations D1 and D2. The pseudo-spin vectors $\mathbf{u_{\mp,k}}$ point in opposite directions and exhibit the same sense of rotation (winding) with quasimomentum $k$. In the phase D1 (D2) $\mathbf{u_{\mp,k}}$ evolve (anti-)\, clockwise and therefore exhibit opposite winding. \label{Fig_1}} 
\end{figure}

In our experiment, the key idea is to combine coherent Bloch oscillations with Ramsey interferometry to determine the geometrical Zak phase and reveal the underlying topological character of the Bloch bands. Previously, the measurement of topological invariants was confined to two-dimensional bands by exploiting the relation between the Chern number and the Hall conductivity for a filled band introduced by Thouless-Kohmoto-Nightingale-DeNijs~\cite{TKNN}. In contrast, in our case the integration over the Brillouin zone necessary for extracting topological invariants is achieved by adiabatic transport of a single-particle wave packet through the band using Bloch oscillations. Recently, it has also been suggested that in the context of ultra-cold atoms, topological properties could be studied through time-of-flight images~\cite{Alba:2011,Zhao:2011,Goldman:2012} or measurements of anomalous velocity~\cite{NiuReview,Price2012}.\\

In the following we focus on a dimerized optical lattice with two sites per unit cell -- a system which, despite its simplicity, exhibits rich topological physics, and depending on the parameter values can mimic either polyacetylene~\cite{SSH}, or conjugated diatomic polymers~\cite{RiceMele}. Within a tight-binding model, the physics of such a system is captured by the Rice-Mele Hamiltonian~\cite{RiceMele}:
\begin{align}
\hat{H}=&-\sum_{n}\left(J\hat{a}^\dagger_{n}\hat{b}_{n}^{\phantom{\dagger}}+J^{\prime}\hat{a}^\dagger_{n}\hat{b}_{n-1}^{\phantom{\dagger}}+\text{h.c.}\right) \nonumber \\
 &+\Delta \sum_n (\hat{a}^\dagger_{n}\hat{a}_{n}-\hat{b}^\dagger_{n}\hat{b}_{n}), \label{eq:ham1}
\end{align}

\noindent where $J$, $J^{\prime}$ denote modulated tunneling amplitudes within the unit cell, $\hat a^{\dagger}_n (\hat b^{\dagger}_n)$ are the particle creation operators for an atom on the sublattice site $a_n (b_n)$ in the $n$th lattice cell (Fig.~\ref{Fig_1}a), and $\Delta$ characterizes the energy offset between neighboring lattice sites.\\

When the on-site energies of the two sites are tuned to be equal ($\Delta=0$), our system corresponds to the SSH model of polyacetylene (see Fig.~\ref{Fig_1}a). In this case, $\hat{H}$ is known to exhibit two topologically distinct phases, D1 for $J>J^{\prime}$, and D2 for $J<J^{\prime}$, separated by a topological phase transition point at $J=J^{\prime}$. The distinct topological character of the two phases is reflected in the difference of their Zak phases, for which $\delta\varphi_\text{Zak}=\pi$. At half filling, a domain wall between phases D1 and D2 features fractionalized excitations. When the on-site energies are tuned to be different ($\Delta \neq 0$), our system models a linearly conjugated diatomic polymer; in this case, the difference of the Zak phases is fractional in units of $\pi$, which generally gives rise to irrationally charged domain walls~\cite{RiceMele}.\\

In the experiment, we realized the Hamiltonian $\hat H$ of Eq.~(\ref{eq:ham1}) by loading a Bose-Einstein condensate of $^{87}$Rb into a one-dimensional optical superlattice potential~\cite{Foelling2007}. This potential was formed by superimposing two standing optical waves of wavelengths $\lambda_s=767$\,nm and $\lambda_l=2\lambda_s=1534$\,nm that generate a lattice potential of the form $V(x)=V_l $sin$^2(k_l x+\phi/2)+V_s $sin$^2(2k_lx + \pi/2)$, where $k_l=2\pi/\lambda_l$ (Fig.~\ref{Fig_1}a). Phase control between the two standing wave fields enabled us to fully control $\phi$. For example, switching between $\phi=0$ and $\phi=\pi$ allowed us to rapidly access the two different dimerized configurations D1 (D2) with $\Delta=0$ in the experiment, whereas by tuning $\phi$ slightly away from these symmetry points, we could introduce a controlled energy offset $\Delta$.\\

The eigenstates of $\hat H$ can be written as Bloch waves of the form:
\begin{align}
\psi_k(x)= e^{ikx} u_k(x) =& \sum_n \alpha_k e^{ikx_n} w_a(x-x_n) \nonumber \\  \nonumber
&+\beta_k e^{ik(x_n+d/2)} w_b(x-x_{n}-d/2), \label{eq:ansatz}
\end{align}
\noindent where $k$ denotes the quasimomentum, $x_n= n\,d$ with $n$ integer, and $w_{a,b}(x)$ are the Wannier functions~\cite{Wannier} for $a_n,b_n$ sites, respectively. The coefficients $\alpha_k, \beta_k$ are determined through the eigenvalue equation $\hat H \psi_k = E_k  \psi_k$. In this case, the cell-periodic wave function $u_k$ can be viewed as a two-component spinor ${\bf u}_{k}=(\alpha_k, \, \beta_k)$, and Eq.~(\ref{eq:Zak}) for the Zak phase takes an especially simple form: 
\begin{equation}
\varphi_\text{Zak}={i}\int_{-G/2}^{G/2} \left( \alpha_k^* \partial_k \alpha_k+\beta_k^* \partial_k \beta_k \right)\, dk.\nonumber
\end{equation}

\begin{figure}
\begin{center}
\includegraphics[width=0.95\linewidth]{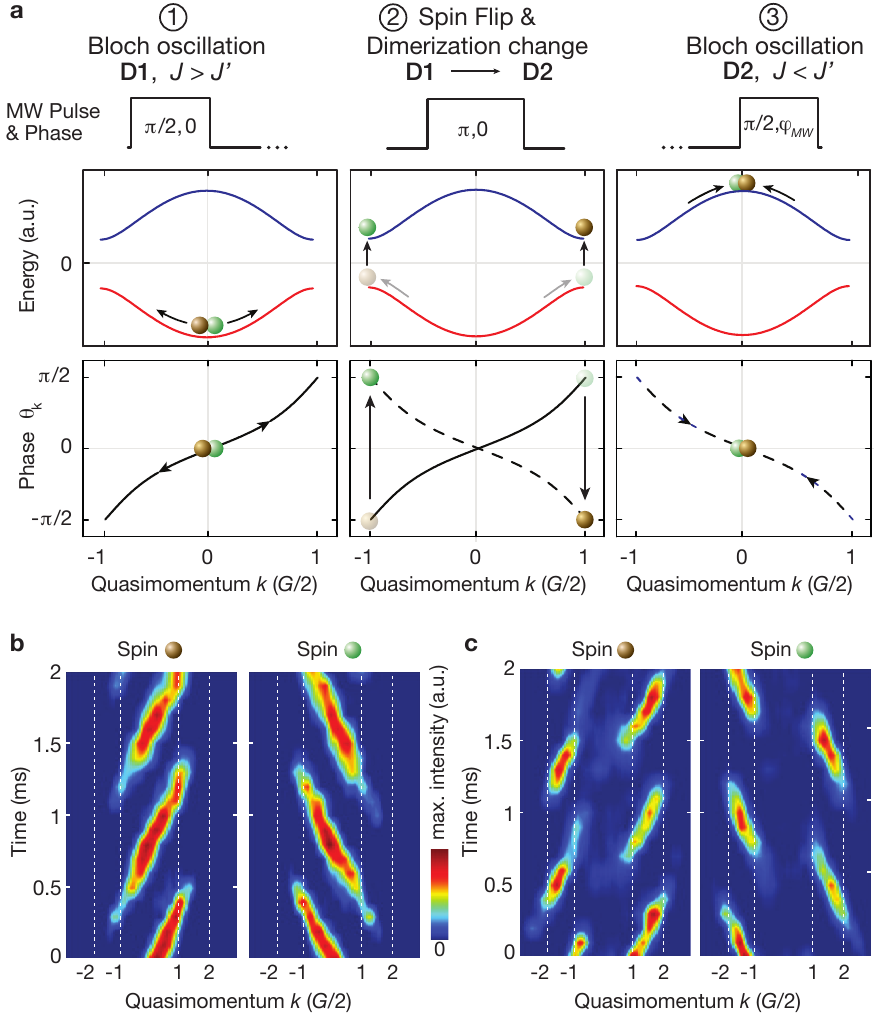}
\end{center}
\caption{Experimental sequence and spin-dependent Bloch oscillations. {\bf (a)} Energy band, MW pulses and state evolution of a single atom in a superposition of two spin-states with opposite magnetic moment (brown and green balls) during the three-step echo sequence described in the text. The winding of the state vector with $k$ is given by $\theta_k$ (solid line dimerization D1, dashed line dimerization D2). {\bf (b,c)} Time-of-flight momentum distributions taken for different evolution times of the spin-dependent Bloch oscillations in the lower (b) and upper energy band (c) used in the experiment. Each momentum point is an average of three identical measurements.}
\label{Fig_2}
\end{figure}

For our choice of the unit cell~\cite{Suplements}, the eigenfunctions for the lower (upper) band of the SSH model ($\Delta=0$) are 
\begin{equation}
\mathbf{u}_{\mp,k} =\frac{1}{\sqrt{2}} \begin{pmatrix} \pm 1 \\  e^{-i \theta _k} \end{pmatrix} ,\nonumber
\end{equation}
\noindent where $\theta_k$ is determined through $J e^{i k d/2}+J^{\prime} e^{-i k d/2}=|\epsilon_k|e^{i \theta_k}$ (see \si). We can thus visualize the Bloch periodic functions as pseudo spin-1/2 states oriented in the equatorial plane of a Bloch sphere (Fig.~\ref{Fig_1}c). Note also that although $\psi_{k+G}(x)=\psi_k(x)$, this translational invariance is not true for $\mathbf u_{\mp,k}$, because in our system with a two-site unit cell $\mathbf u_{\mp,k+G}=  \hat \sigma_{z} \mathbf u_{\mp,k}$, where $\hat\sigma_z$ is the third Pauli matrix. As the two state vectors for the upper and lower bands are orthogonal, they point in opposite directions and therefore exhibit the same winding when the quasimomentum $k$ is varied adiabatically. The Zak phases for the lower and upper band are thus identical $\varphi^{D1}_\text{Zak} = \pi/2$. However, when the dimerization is changed from configuration D1 to D2 (Fig.~\ref{Fig_1}c), the corresponding geometric phase changes to $\varphi^{D2}_\text{Zak}=-\pi/2$, because of the opposite winding of the state with quasimomentum $k$. The difference of the two Zak phases for the two dimerized configurations is then:
\begin{equation}
  \delta \varphi_\text{Zak} = \varphi^{D1}_\text{Zak} - \varphi^{D2}_\text{Zak} = \pi.
\end{equation}
We point out that the Zak phase of each dimerization is a gauge dependent quantity, i.e.\ it depends on the choice of origin of the unit cell, however, the difference of Zak phases of the two dimerizations is uniquely defined~\cite{Vanderbilt,Qi:2011}. 

 When an atom is adiabatically evolved through the Brillouin zone of the periodic potential $k\rightarrow k+G$, it acquires a phase shift due to three distinct contributions: i) a geometric phase $\varphi_\text{Zak}$ as well as ii) a  dynamical phase $\varphi_\text{dyn}=\int E(t)/\hbar \, dt$, both derived from the band-structure, and iii) a phase due to the Zeeman energy of the atom in an external magnetic field (see semiclassical analysis in \si): 
\begin{equation}
  \varphi_\text{tot} = \varphi_\text{Zak} + \varphi_\text{dyn} + \varphi_\text{Zeeman}.\nonumber
\end{equation}

To isolate the geometrical Zak phase in the experiment, we employ a three-step sequence (Fig.~\ref{Fig_2}a and \si). Step 1) We start with an atom in the state $|\!\!\downarrow,k=0\rangle$ and bring it into a coherent superposition state $1/\sqrt{2}(|\!\! \uparrow,k=0\ra+|\!\!\downarrow,k=0\ra)$ using a microwave $\pi/2$-pulse. Here $\sigma=\uparrow,\downarrow$ denote two spin states of the atom with opposite magnetic moment. Then a magnetic field gradient is applied that creates a constant force in opposite directions for the two spin components. Such a constant force leads to Bloch oscillations, i.e. a linear evolution of quasimomentum over time~\cite{Salomon96}. In our case the force is directed in opposite directions for the two spin components. The atomic wavepacket thus evolves into the coherent superposition state $1/\sqrt{2}(|\!\! \uparrow,k\ra+e^{i \delta \varphi}|\!\!\downarrow,-k\ra)$. When both reach the band edge, the differential phase between the two states is given by $\delta \varphi = \varphi_\text{Zak}+\delta \varphi_\text{Zeeman}$. Note that the dynamical phase acquired during the adiabatic evolution is equal for the two spin states and therefore cancels in the phase difference. Step 2) To  eliminate the Zeeman phase difference, we apply a spin-echo $\pi$-pulse at this point and also switch dimerization from D1$\rightarrow$D2. For atoms located at the band edge $k=\pm G/2$, this non-adiabatic dimerization switch induces a transition to the excited band of the SSH model. Step 3) The sequence is finally completed by letting the spin components further evolve in the upper band until they return to $k=0$. At this point in time, a final $\pi/2$-pulse with phase $\varphi_{MW}$ is applied in order to interfere the two spin components and read out their relative phase $\delta \varphi$ through the resulting Ramsey fringe. The change in dimerization occurring at the mid-point of the echo sequence is crucial in order not to cancel the Zak phase in addition to the Zeeman phase. Due to the opposite windings of the Bloch states in the upper and lower bands with quasimomentum $k$ (Fig.~\ref{Fig_1}c), the resulting phase shift encoded in the Ramsey fringe is thus given by: $\delta \varphi = \varphi^{D1}_\text{Zak} - \varphi^{D2}_\text{Zak}$ if the dimerization is swapped, whereas $\delta \varphi=0$ if it is left unchanged. \\

In Fig.~\ref{Fig_2}b,c we show images of the momentum distribution of the atoms during the spin-dependent Bloch oscillations in the lower and upper energy bands. Note the opposite evolution in momentum space due to the opposite magnetic moments of the two spin-states. Atoms in the upper energy band are characterized by a distinctively different momentum pattern from atoms in the lower energy band. The Bloch oscillations period of $\tau_\text{Bloch}=0.85(3)\,$ms was chosen to be slow enough, such that non-adiabatic Landau-Zener transitions at the band edge are negligible, while still maintaining an overall fast evolution time to minimize decoherence effects.\\

\begin{figure}
\begin{center}
\includegraphics[width=0.95\linewidth]{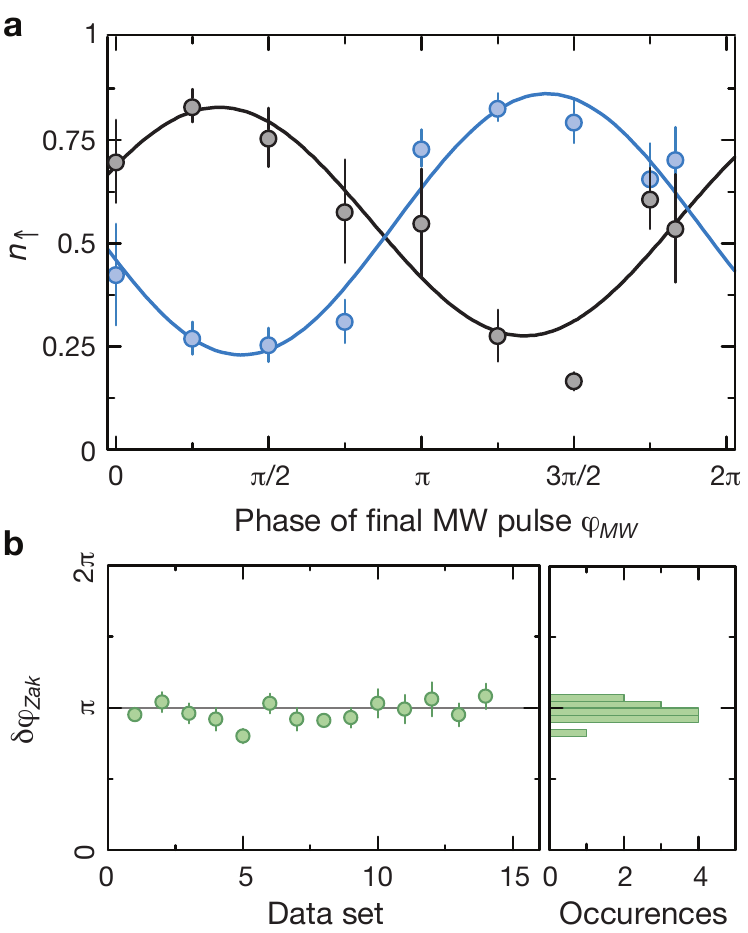}
\end{center}
\vspace{-0.5cm} \caption{Determination of the Zak phase. {\bf (a)} Following the sequence described in the text, the atom number in the two spin states $N_{\uparrow,\downarrow}$ is measured and the fraction of atoms in the $|\!\!\uparrow\rangle$ spin state $n_\uparrow=N_\uparrow/(N_\uparrow+N_\downarrow)$ is plotted as a function of the phase of the final microwave $\pi/2$-pulse. The difference in phase of the two Ramsey fringes yields the Zak phase difference $\delta \varphi_\text{Zak} = \varphi_\text{Zak}^{D1} - \varphi_\text{Zak}^{D2}$. Blue (black) circles correspond to the fringe in which the dimerization was (not) swapped. In order to reduce the effect of fluctuations, every data point is an average of five individual measurements and the error bars show the standard deviation of the mean. The phase of the reference fringe (black) is determined by a small detuning of the microwave pulse~\cite{Suplements}. {\bf (b)} Measured relative phase for 14 identical experimental runs (left), which give an average value of $\delta\varphi_\text{Zak}=0.97(2)\pi$. The corresponding histogram is shown on the right with a binning of $0.05\pi$. The $1\sigma$-width of the resulting distribution is $\sigma=0.07\pi$. \label{Fig_3}} 
\end{figure}

A typical result for the two Ramsey fringes obtained with and without dimerization swapping during the state evolution can be seen in Fig.~\ref{Fig_3}a. Each plotted value for a given angle $\varphi_{MW}$ is an average over five identical measurements in order to reduce the effect of residual fluctuations. We performed a further statistical analysis by recording 14 independent Ramsey fringes for the two configurations. The obtained phase differences are shown in Fig.~\ref{Fig_3}b together with the corresponding histogram. From these individual measurements we determine the geometric phase difference between the two dimerized configurations to be: 
\begin{equation}
  \delta \varphi = 0.97(2) \pi,\nonumber
\end{equation}

\noindent in excellent agreement with theory, as discussed above. The uncertainty in the recorded value denotes the standard error of the mean obtained from the distribution function (Fig.~\ref{Fig_3}b) and is mainly determined by experimental imperfections in the control of the underlying lattice potentials.\\

\begin{figure}
\begin{center}
\includegraphics[width=0.95\linewidth]{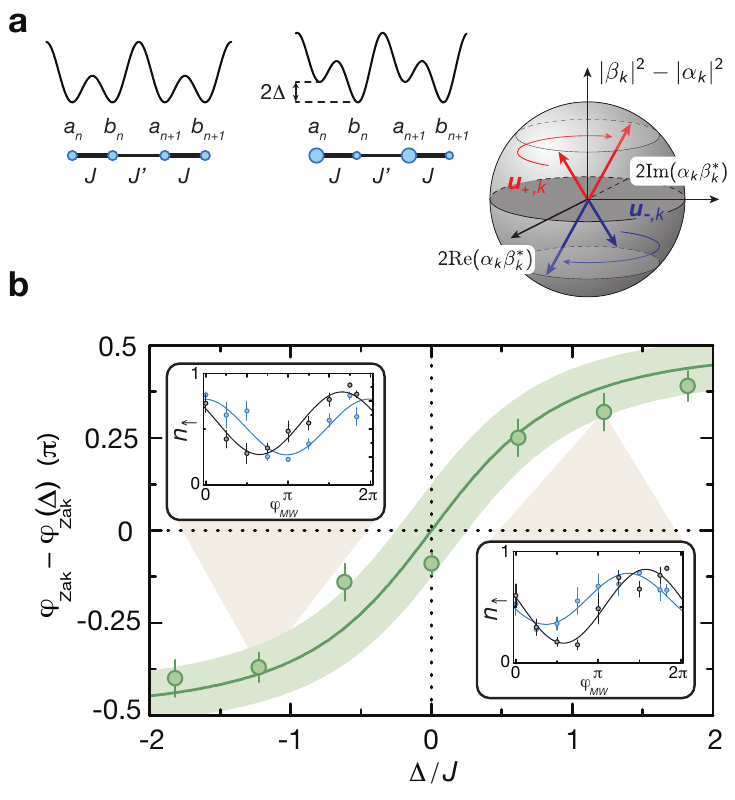}
\end{center}
\vspace{-0.5cm} \caption{Fractional Zak phase. {\bf (a)} Lattice potential without and with an on-site energy staggering $\Delta$. When $\Delta=0$ the Zak phase is $\varphi_\text{Zak}(\Delta=0)=\pi/2$. As $\Delta$ increases, the pseudo-spin vectors move away from the equatorial plane and the value of $\varphi_\text{Zak}(\Delta)$ decays rapidly to zero. {\bf (b)} Measured phase difference $\varphi_\text{Zak}-\varphi_\text{Zak}(\Delta)$ as a function of $\Delta$. Each individual point was obtained from four individual measurements. The vertical error bars represent the standard error of the mean. The green line is the theoretical prediction and the shaded area represents the uncertainties in the calibration of the energy offset $\Delta$. The insets show a typical Ramsey fringe for $\Delta/J=-1.2$ (left) and $\Delta/J=1.2$ (right), which were used to extract the relative phase $\delta \varphi$. The blue (black) fringes correspond to measurements with (without) staggering~\cite{Suplements}. \label{Fig_4}} 
\end{figure}

To further investigate the variation of the Zak phase with lattice parameters, we performed a second series of experiments with a staggered on-site energy offset $\Delta$ (Fig.~\ref{Fig_4}a), corresponding to a heteropolar dimer configuration~\cite{RiceMele}. The energy offset $\Delta$ displaces the pseudo-spin Bloch vectors away from the equatorial plane, resulting in an additional dependence of the Zak phase on the offset $\Delta$  (Fig.~\ref{Fig_4}a and \si). In order to probe the dependence of $\varphi_\text{Zak}$ on $\Delta$, we performed an experimental sequence that was similar to the one described above. However, instead of swapping the dimerization from D1 to D2, an energy offset $|\Delta| < 2J$ was introduced for one half of the sequence. Thereafter, because of the spin-echo pulse, the wavepackets return to $k=0$ in the lowest band. Though the system completes a full Bloch oscillation in the lowest band, the total geometric phase acquired is not zero, since the Bloch vector is displaced from the equatorial plane during one half of the sequence (Fig.~\ref{Fig_4}) and the Zak phase is changed from $\varphi_\text{Zak}$ to $\varphi_\text{Zak}(\Delta)$. The resulting phase in the Ramsey fringe is thus given by $\delta \varphi=\varphi_\text{Zak}-\varphi_\text{Zak}(\Delta)$ when the energy offset is present, and it is $\delta \varphi=0$ when the offset is absent. As before, the phase difference between these two fringes for atoms in the lowest band allows us to determine the relative phase $\varphi_\text{Zak}-\varphi_\text{Zak}(\Delta)$. During the non-adiabatic switching of the superlattice potential at step 2 of the experimental sequence  some of the atoms are transferred to the higher band and acquire a different geometric phase. However, taking into account this contribution to the measured phase difference enabled us to extract the relative phase $\delta \varphi$ from our data (see \si). As shown in Fig.~\ref{Fig_4}b, we find good agreement between the measured and predicted values of the fractional Zak phase.\\

In conclusion, we have presented a general approach for studying topological properties of Bloch bands in optical lattices and demonstrated its versatility through a first direct measurement of the topological invariant in topologically non-trivial Bloch bands. Making use of the recently demonstrated control of optical potentials at the single-site level~\cite{Weitenberg:2011}, we plan to realize domain walls in the dimerized lattice that would allow us to directly study edge states~\cite{Kitagawa:2012,Kraus:2012} and fractional charges located at the interface of the two topologically distinct phases~\cite{Jackiw76,Wilczek,SSH,Ruostekoski:2002}. Although in this work we focused on one-dimensional systems, our technique can easily be extended to two-dimensional systems, where the change of the Zak phase in the Brillouin zone gives the topological density of the Bloch band~\cite{Dima}. This enables measurements of both the Chern number of topological bands and the $\pi$-flux associated with a Dirac point. Additionally, we expect that this idea can be extended to measure the non-Abelian Berry's phase in Bloch bands, such as in a system with quantum spin Hall effect~\cite{Kane:2005}, to the study of Floquet states in periodically driven systems~\cite{Grifoni:1998,Kitagawa:2010,Lindner:2011}, and to quasiparticles in unconventional superconductors, such as $d$-wave superconductors, which have Dirac dispersion at the nodal points~\cite{Volovik:2003}. Overall, our work indicates that cold atomic systems provide a versatile platform for studying topological states of matter, and establishes a novel method for probing their properties.\\

We acknowledge helpful discussions with B. Paredes. This work was supported by the DFG (FOR635, FOR801) and DARPA (OLE program). M. Ai. was additionally supported by the Deutsche Telekom Stiftung.

\section{\si}

\bigskip
\renewcommand{\theequation}{S\arabic{equation}}
\setcounter{equation}{0}
 \renewcommand{\thesection}{S.\Roman{section}}

\section{Discussion of the Zak phase in the dimerized lattice}

In this section we discuss the Bloch wave functions in the SSH model, and calculate the Zak phase of two dimerized configurations of the model.  We consider the most general case when not only tunneling
but also on-site potentials on sublattices A and B are dimerized~\cite{RiceMele2}:
\begin{align}
\hat{H}=&-\sum_{n}\left(J\hat{a}^\dagger_{n}\hat{b}_{n}^{\phantom{\dagger}}+J^{\prime}\hat{a}^\dagger_{n}\hat{b}_{n-1}^{\phantom{\dagger}}+\text{h.c.}\right) \\ \nonumber
 &+\Delta \sum_n (\hat{a}^\dagger_{n}\hat{a}_{n}-\hat{b}^\dagger_{n}\hat{b}_{n}).
\label{Hdim}
\end{align}

\noindent The energy bands and eigenstates can be obtained from the eigenvalue equation
\begin{eqnarray}\label{eigenstates}
\left[ \begin{array}{cc} \Delta & -\rho_k \\
-\rho_k^* & -\Delta \end{array} \right] \left( \begin{array}{c} \alpha_k \\ \beta_k \end{array} \right) = 
\tilde{\epsilon}_k \left( \begin{array}{c} \alpha_k \\ \beta_k \end{array} \right) ,
\end{eqnarray}
where 
\be\label{rho}
\rho_k=Je^{ikd/2}+J'e^{-ikd/2}=|\epsilon_k| e^{i\theta_k}. 
\ee
The variable $\epsilon_k$ in the above expression corresponds to the energy of Bloch states in the SSH model ($\Delta=0$).  

\subsection{Eigenstates}

Solving Eq.~(\ref{eigenstates}), we obtain for the eigenenergies 
\be\label{eq:tilde_e}
\tilde{\epsilon}_k = \pm \sqrt{\Delta^2 + \epsilon_k^2}.
\ee
Following Ref.\,\cite{Zak892}, we require that  $\alpha_{k+G} = \alpha_k$ and $\beta_{k+G}= -\beta_k$.
The eigenstates in the lower and upper bands which satisfy this condition can be chosen as follows: 
\begin{eqnarray}
\left( \begin{array}{c} \alpha_{-,k} \\ \beta_{-,k} \end{array} \right)
&=& \left( \begin{array}{c} \sin \frac{\gamma_k}{2} \\ \cos \frac{\gamma_k}{2} e^{-i \theta_k} \end{array} \right)
\nonumber\\
\left( \begin{array}{c} \alpha_{+,k} \\ \beta_{+,k} \end{array} \right)
&=& \left( \begin{array}{c} -\cos \frac{\gamma_k}{2}  \\ \sin \frac{\gamma_k}{2} e^{-i\theta_k}  \end{array} \right),
\label{Wavefunctions}
\end{eqnarray}
where $\gamma_k$ and $\theta_k$ are given by (the expression for $\theta_k$ is obtained from Eq.~(\ref{rho}))
\begin{eqnarray}
\gamma_k &=& \arctan \frac{\epsilon_k}{\Delta} \,
\nonumber\\
\theta_k &=& \arctan \frac{(J-J')\sin (kd/2)}{(J+J')\cos (kd/2)}. 
\nonumber
\end{eqnarray}

\subsection{Zak phase}

Generally, the Zak phase for a given band is given by
\be\label{eq:zak_alpha_beta}
\varphi_{\rm Zak}=i\int_{-G/2}^{G/2} (\alpha_k^* \partial_k \alpha_k+\beta_k^* \partial_k \beta_k) dk. 
\ee
For the wave functions in Eq.~(\ref{Wavefunctions}), this gives
\begin{align}\label{eq:zak2}
\varphi_{\rm Zak,-}(\Delta)=&\int_{-G/2}^{G/2} \cos^2 \frac{\gamma_k}{2} \partial_k\theta_k \, dk,\\ \nonumber
\varphi_{\rm Zak,+}(\Delta)=&\int_{-G/2}^{G/2} \sin^2 \frac{\gamma_k}{2} \partial_k\theta_k \, dk. 
\end{align}
These equations were used to make the green theoretical curve in Fig.~4 of the main text.\\ 

For the SSH model $\Delta=0$ such that $\cos^2\gamma_k/2=\sin^2\gamma_k/2=1/2$ and the Zak phases reduce to $\varphi^{\rm D1}_{\rm Zak,-}=\frac{\pi}{2}$ and $\varphi^{\rm D2}_{\rm Zak,+}=- \frac{\pi}{2}$.

\section{Zak phase and the choice of a unit cell}

We emphasize that the Zak phase, in general, depends on the choice of unit cell for the crystal. Here we would like to clarify why for our choice of the unit cell of the SSH model the Zak phases are given by $\pm \pi/2$ (rather than by $0$ or $\pi$, as in some previous papers, e.g., see Ref.~\cite{Montambaux2} and references therein).\\ 

To understand the relation of the Zak phase to the choice of the unit cell, we first note that the function $u_k(x)$ is not periodic under a translation by a reciprocal lattice vector $G=2\pi/d$, but rather satisfies the following condition (see Ref.~\cite{Zak892}): 
\be\label{eq:u-periodicity}
u_{k+G}(x)=e^{-iGx} u_k(x).  
\ee
This guarantees that the full Bloch function $\psi_k(x)=e^{ikx} u_k(x)$ is periodic under a translation by a reciprocal lattice vector $G$.\\ 

The Zak phase is defined in terms of the periodic Bloch functions (see Ref.\,\cite{Zak892}): 
\be\label{eq:zak_phase}
\varphi_\text{Zak}=\int_{-G/2}^{G/2}  A(k)\, dk, \,\, A(k)=i\la u_k | \partial_k u_k\ra,
\ee
where $A(k)$ is the Berry's connection, and the scalar product is defined as $\la u | v\ra=\int _0^d u^*(x) v(x) dx $.\\ 

Let us now consider a translation by a distance $a$: $x'=x+a$, which leads to a different choice of the unit cell (unless $a=nd$, where $n$ is integer). The new unit cell is $x'\in [0,d)$, or, equivalently, $x\in [-a,d-a)$. Then, in order for the condition in Eq.\,(\ref{eq:u-periodicity}) to be satisfied, the Bloch functions have to be redefined as follows: 
\be\label{eq:u'}
u'_k(x')=u_k(x'-d)e^{-ika}. 
\ee
The Berry connection is modified accordingly:
$$
A'(k)=\la u'_k| \partial_k u'_k \ra=
\la u_k| \partial_k u_k \ra-ia=A(k)-ia.  
$$

Integrating the Berry connection over the Brillouin zone, we obtain the change in the Zak phase:
\be\label{eq:Zak_change}
\varphi_{\rm Zak}'=\varphi_{\rm Zak}+Ga=\varphi_Z+2\pi a/d. 
\ee
The Zak phase changes by $2\pi$ under a translation by a multiple of the lattice vector $d$. Thus it is not surprising that our choice of the unit cell for the SSH model gives rise to the values of the Zak phase which are different from the values mentioned in some other works~\cite{Montambaux2}. The difference of $\pi/2$ appears due to the shift of the unit cell by $a=d/4$.\\ 

We also emphasize that, even though the Zak phase itself depends on the choice of the unit cell, the difference of the Zak phases in two states, measured in our experiment, is an invariant which describes the topological properties of the two states (see Ref.~\cite{Vanderbilt2}). 

\section{Bloch oscillations in superlattices}

Here we derive the dynamical equations for the Bloch oscillations in the dimerized lattice. We consider a dimerized lattice subject to an external force, described by the Hamiltonian 
\begin{align}
{\hat H_F} = {\hat H} - F \sum_n \left\{ \, (x_n-x_0) \hat a_n^\dagger \hat a_n + (x_n+\frac{d}{2}-x_0) \hat b_n^\dagger \hat b_n\,\right\}
\nonumber
\end{align}
with $x_n = nd$. For simplicity, we wrote this equation for only one of the pseudo-spin species; an equation for the dynamics of other pseudo-spin is obtained by a substitution $F\to -F$ in the above equation. Slowly fluctuating magnetic fields in our experiment will add an extra Zeeman energy difference between two pseudo-spins, which can be absorbed in the definition of $x_0$.\\ 

Eigenstates of ${\hat H_F}$ are plane waves with energies $\pm \tilde\epsilon_{k}=\pm \sqrt{\Delta^2+\epsilon_k^2}$. For our analysis it is convenient to work in second quantization, 
\begin{align}
\hat c^\dagger_{\pm,k} =& \frac{1}{\sqrt{N}} \sum_n \Bigr[
  \alpha_{\pm,k}e^{ikx_n} \hat a_n^\dagger \\ \nonumber
+& \beta_{\pm,k}e^{ik(x_n+\frac{d}{2})} \hat b_n^\dagger \Bigr]
\end{align}
where $N$ is the total number of sites in the lattice. For finite $F$ we need to solve the Heisenberg equation of motion $\frac{d}{dt} \hat\Psi^\dagger(t) = \frac{i}{\hbar}[{\hat H_F}, \hat \Psi^\dagger(t)]$.
We look for solutions where the quasimomentum changes at a constant rate, 
\begin{eqnarray}
\hat \Psi^\dagger(t) = A(t) \hat c^\dagger_{-,k_0-vt } + B(t) \hat c^\dagger_{+,k_0-vt }. 
\label{BlochAnsatz}
\end{eqnarray}
Using 
\begin{align}
&\frac{\partial}{\partial t} \hat c^\dagger_{-,k_0-vt} =\nonumber \\
&- \frac{v}{\sqrt{N}}  \sum_n   \left [ i \alpha_{-,k} x_n + \frac{\partial \alpha_{-,k}}{\partial k} \right ]e^{ik x_n}  \hat a_n^\dagger \,\,|_{k=k_0-vt}
\nonumber\\
&-\frac{v}{\sqrt{N}} \sum_n \left [ i \beta_{-,k}(x_n + \frac{d}{2}) \right.\nonumber \\
&+\left. \frac{\partial \beta_{-,k}}{\partial k} \right ] e^{ik(x_n+\frac{1}{2})}  \hat b_n^\dagger \,\, |_{k=k_0-vt}
\end{align}
and
\begin{align}
[ {\hat H}, \hat c^\dagger_{-,q} ] =&  \epsilon_{-,q} \hat c^\dagger_{-,q } 
- \frac{F}{\sqrt{N}} \sum_n \left [  \alpha_{-,q} (x_n-x_0) e^{iqx_n} \hat a_n^\dagger  \frac{}{}\right.
\nonumber\\
&+ \left. \beta_{-,q} (x_n+\frac{d}{2}-x_0) e^{iq(x_n+\frac{d}{2})} \hat b_n^\dagger \right ],
\nonumber
\end{align}
we find that the Ansatz in Eq.~(\ref{BlochAnsatz}) provides a solution of the Heisenberg equation of motion
when $v=f=F/\hbar$ and
\begin{align}
-i \dot{A} = & \epsilon_{-,k_0-ft} A + f x_0 A - \frac{fA}{i}\langle u_{-,k_0-ft} | \partial_k u_{-,k_0-ft} \rangle 
\nonumber\\
&- \frac{fB}{i}\langle u_{-,k_0-ft} | \partial_k u_{+,k_0-ft} \rangle
\label{BlochEOM}
\end{align}
and a similar equation for $B$. The last term in Eq.~(\ref{BlochEOM}) describes non-adiabatic mixing of the bands, which we neglect. Assuming that atoms occupy only the lower band ($B=0$ and $|A|=1$) and taking $A(t)=e^{i\varphi(t)}$, we obtain:
\begin{align}
\dot{\varphi} = \epsilon_{-,k_0-f t} +f x_0 - \frac{f}{i}\langle u_{-,k_0-ft} | \partial_k u_{-,k_0-ft} \rangle.
\label{BlochEOM2}
\end{align}
The first term in Eq.~(\ref{BlochEOM2}) describes the dynamical phase contribution, the second is, effectively, the Zeeman phase (which contains the effect of the fluctuating magnetic fields that shift $x_0$), and the last term describes the Berry's phase part.  Integrating Eq.~(\ref{BlochEOM2}) over a period of the Bloch oscillations gives Eq. (3) of the main text, with $\varphi_{\rm Zeeman}=\pm fx_0 t$, where $\pm$ sign corresponds to the two pseudo-spin species.\\ 

A crucial point is that our spin-echo type protocol is insensitive to both the Zeeman phase (or, equivalently, to the value of $x_0$) and the dynamical phase. The spin-echo nature of the experiment guarantees that the Zeeman phases for the two trajectories are equal (each atom spends half of the time in the spin-up state, and half of the time in the spin-down state). The dynamical phases acquired during the motion in the Bloch bands cancel out due to the reflection symmetry of the dispersion relation, $\epsilon_k=\epsilon_{-k}$. Both of these points finally allow us to single out the Zak phase.

\section{Experimental Sequence}

The experimental sequence started by loading in $200\,$ms a Bose-Einstein condensate of about $5\times 10^4$ atoms in the Zeeman state $\left|F=1, m_F=-1\right>=\left|\downarrow\right>$ into a dimerized 1D lattice of $V_l=13E_{r,l}$, $V_s=4E_{r,s}$ and $\phi=0$, where $E_{r,i}=h^2/2m\lambda_i$, $i=l,s$. Thereafter, a magnetic field gradient of $f=1.18(4)$~kHz$/d$ was ramped up in $1.5$~ms, and subsequently a $3$-$\mu$s MW $\pi/2$-pulse coupling to the Zeeman state $\left|F=2, m_F=-1\right>=\left|\uparrow\right>$ was applied. After half a Bloch oscillation, when the atoms were at the edge of the band, a $\pi$-pulse was applied and the dimerization was swapped (See section Dimerization swap). After an additional time of $425\,\mu$s the atoms returned to $k=0$ and a final $\pi/2$-pulse with a phase $\varphi_{MW}$ was applied. At the end of the sequence we measured the population fraction of each of the two spin components by using time-of-flight imaging with a Stern-Gerlach gradient field applied during the expansion to separate the different Zeeman states.\\

For the $850$-$\mu$s Bloch oscillation time and $J/J^{\prime}=1~$kHz$/0.06$~kHz, a negligible fraction of atoms was transferred to the upper band because of non-adiabatic Landau-Zener transitions at the band edge. This underlines the adiabaticity of the Bloch oscillations.\\

Due to a reproducible $80$-kHz drift in the offset magnetic field during the time of a full Bloch oscillation, the three MW pulses were not all on resonance. The first (last) $\pi/2$-pulse was detuned by $-40$~kHz ($40$~kHz) and the intermediate $\pi$-pulse was on resonance, while the coupling strength is about $80$~kHz. The drift in the offset magnetic field adds a constant value to the Zeeman phase $\varphi_\text{Zeeman}$, which is independent of the change in dimerization/staggering, therefore it does not affect the phase difference $\delta \varphi$.

\section{Dimerization swap}

In order to change the dimerization from $D1$ to $D2$ we employed a second long lattice, whose amplitude and relative phase $\phi^{\prime}$ can be independently controlled. The dimerization exchange was performed by quickly switching off the first long lattice with phase $\phi=0$ and a ramp-up of the second long lattice with a phase $\phi^{\prime}=\pi$ within $10~\mu$s. As the dimerization swap time is much shorter than the Bloch oscillation time, the dimerization exchange can be considered instantaneous.
The resulting lattice potentials of the different lattice configurations are shown in Fig.~\ref{Fig_5}, corresponding to the two dimerizations of the SSH model.
\vspace{0.5cm}

\setcounter{figure}{0}
\renewcommand{\thefigure}{S\arabic{figure}}
\begin{figure}
\begin{center}\includegraphics[width=0.95\linewidth]{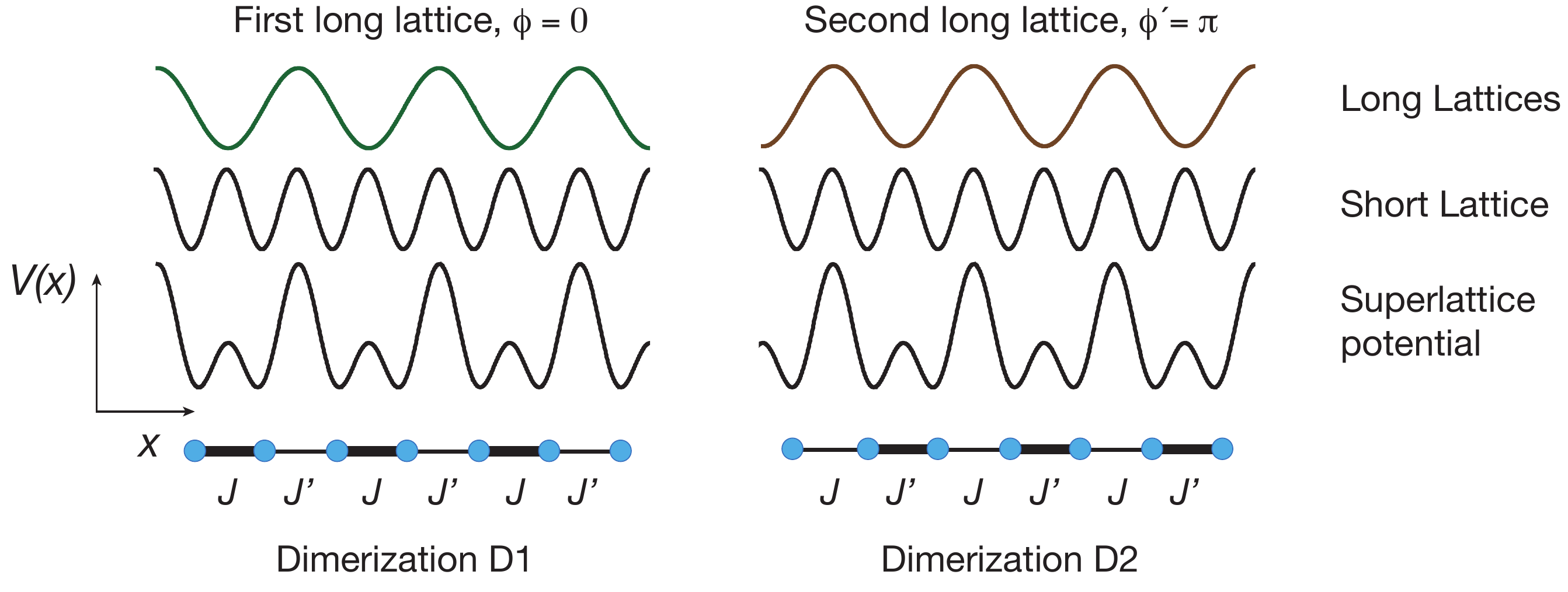}\end{center}
\vspace{-0.5cm} \caption{Left (right): Superlattice potential created by using
  the first (second) long lattice with a phase $\phi=0$ ($\phi^{\prime}=\pi$). The corresponding
  dimerization is $D1$ ($D2$). \label{Fig_5}}
\end{figure}

\begin{figure}[h]
\begin{center}\includegraphics[width=0.95\linewidth]{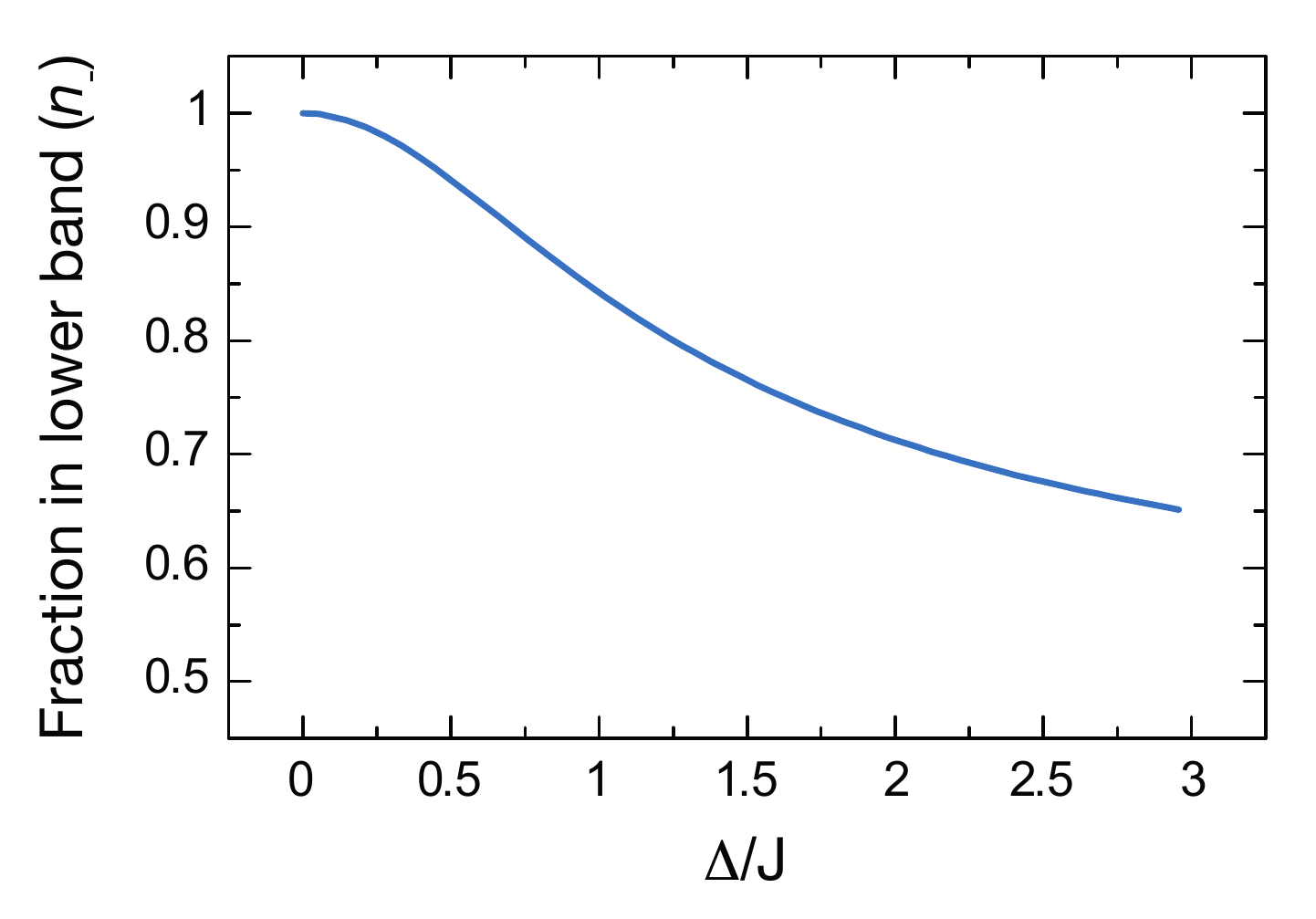}\end{center}
\vspace{-0.5cm} \caption{Atom fraction in the lower band ($n_-$) as a function
  of the energy offset in units of $J$. As $\Delta/J$ increases the fraction transferred to the upper band ($n_+$) also increases. In the limit $\Delta/J \rightarrow \infty, n_- \rightarrow 0.5$  \label{Fig_6}}
\end{figure}

\section{Measuring the Zak phase at $\Delta\neq 0$: projection onto lower and upper bands}

Our protocol for measuring the Zak phase at $\Delta\neq 0$, described in the main text, involves a sudden turn on/off of $\Delta$ when the atoms, initially in the lower band, reach the edges of the Brillouin zone at $k=\pm \pi/d$. In such a process, a fraction of atoms get excited to the upper band. The upper- and lower-band populations $n_{\pm}$ after such a non-adiabatic turn on/off of the staggering $\Delta$ can be obtained by projecting the lower-band eigenstates in Eq.~(\ref{Wavefunctions}) with $\Delta=0$ onto the states with $\Delta\neq 0$. They are given by
\be\label{eq:fraction}
n_{\pm}=\frac{1\mp \sin\gamma_k}{2}=\frac{\tilde{\epsilon}_k\mp \epsilon_k}{2\tilde\epsilon_k}.  
\ee In the above expression, all quantities are defined for the final non-zero value of $\Delta$.

\section{Extracting the phase difference $\varphi_{\mathrm{Zak}}-\varphi_{\mathrm{Zak}}(\Delta)$ from the measured Ramsey fringes}
In this section we explain how we extract the value of $\varphi_{\mathrm{Zak}}-\varphi_{\mathrm{Zak}}(\Delta)$ from the measured fringes for Fig.\,4 of the main text. Due to a fraction of atoms transferred to the upper band when $\Delta\neq 0$, the phase difference between the two Ramsey fringes deviates from $\varphi_{\mathrm{Zak}}-\varphi_{\mathrm{Zak}}(\Delta)$. The reason for this is that the fraction of atoms in the upper band acquires a different geometric phase $\varphi_{\mathrm{Zak},+}(\Delta)=\pi-\varphi_{\mathrm{Zak},-}(\Delta)$ (see Eq.~(\ref{eq:zak2})) than the fraction in the lower band. Therefore, when measuring the Ramsey fringe, both phases enter into play as explained below.\\  

As described in the main text, for half of the measured points during the first part of the Bloch oscillation, the atoms evolve in the lower band of the dimerized system with $\Delta=0$ (for the second half of the measured points the atoms start in the lower band with $\Delta \neq 0$, which gives essentially the same result). When they reach the edge of the band, the wave function is
\begin{align}
 |\psi \ra=\frac{1}{\sqrt{2}}\big(|\!\!\uparrow ,-G/2\ra_-e^{i\varphi_{\mathrm{Zak}}/2}+|\!\!\downarrow,G/2\ra_-e^{-i\varphi_{\mathrm{Zak}}/2}\big),\nonumber
\end{align}

\noindent where $\varphi_{\mathrm{Zak}}=\varphi_{\mathrm{Zak},-}(\Delta=0)$. When the MW $\pi$-pulse is applied and the energy offset $\Delta$ is quickly introduced, the spins are flipped and the population in the bands are $n_{\pm}$ (see Eq.~(\ref{eq:fraction}) and Fig.~\ref{Fig_6}): 
\begin{align} 
|\psi \ra =&  \frac{1}{\sqrt{2}}\Big( \sqrt{n_-} | -G/2 \ra_- + \sqrt{n_+}|-G/2\ra_+ \Big)e^{i\varphi_{\mathrm{Zak}}/2}|\!\!\downarrow \ra
\nonumber \\ 
+&  \frac{1}{\sqrt{2}}\Big( \sqrt{n_-} | G/2 \ra_- + \sqrt{n_+}|G/2\ra_+ \Big)e^{-i\varphi_{\mathrm{Zak}}/2}|\!\!\uparrow \ra. \nonumber
\end{align}

\begin{figure}
\begin{center}\includegraphics[width=0.95\linewidth]{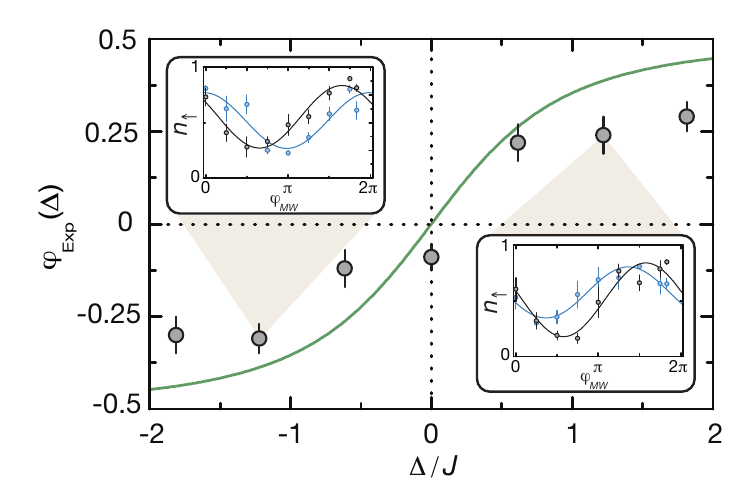}\end{center}
\vspace{-0.5cm} \caption{Measured phase difference $\varphi_{Exp}$ as a function
  of the energy offset in units of $J$. The green line shows the predicted value of $\varphi_{\mathrm{Zak}}-\varphi_{\mathrm{Zak}}(\Delta)$ when there is no population transfer to the upper band.  \label{Fig_7}}
\end{figure}

\noindent In the second half of the Bloch oscillation the particles return to $k=0$ and the fraction in the upper (lower) band picks up a phase $\varphi_{\mathrm{Zak},+}(\Delta)$ ($\varphi_{\mathrm{Zak},-}(\Delta)$), such that the resulting state vector is: 
\vspace{1mm}
\begin{align}
|\psi \ra =&  \frac{1}{\sqrt{2}}\left \{ \sqrt{n_-} |0 \ra_-e^{-i\varphi_{\mathrm{Zak},-}(\Delta)/2} \right.\nonumber\\
+&\left. \sqrt{n_+}|0\ra_+e^{-i\varphi_{\mathrm{Zak},+}(\Delta)/2} \right \}e^{i\varphi_{\mathrm{Zak}}/2}|\!\!\downarrow \ra
\nonumber \\ 
+&  \frac{1}{\sqrt{2}}\left \{ \sqrt{n_-} | 0 \ra_-e^{i\varphi_{Zak,-}(\Delta)/2} \right. \nonumber \\
+& \left.\sqrt{n_+}|0\ra_+e^{i\varphi_{Zak,+}(\Delta)/2} \right \}e^{-i\varphi_\mathrm{Zak}/2}|\!\!\uparrow \ra. \nonumber
\end{align}
\noindent Finally, we apply a $\pi/2$-pulse with a MW phase that rotates the spins to $|\!\!\uparrow\ra \rightarrow  \frac{1}{\sqrt{2}}(e^{-\varphi_{MW}/2}|\!\!\uparrow\ra + e^{+\varphi_{MW}/2}|\!\!\downarrow\ra)$ and $|\!\!\downarrow\ra \rightarrow  \frac{1}{\sqrt{2}}(e^{\varphi_{MW}/2}|\!\!\uparrow \ra + e^{-\varphi_{MW}/2}|\!\!\downarrow \ra)$. After this pulse, the total population in $|\!\!\uparrow \ra$ is given by:
\begin{align}
n_{\uparrow} =& \frac{1}{2}\left \{ 1+ n_{-}\cos(\varphi_{\mathrm{Zak}}-\varphi_{\mathrm{Zak},-}(\Delta)+\varphi_{MW}) \right.\nonumber\\
+& \left. n_{+}\cos(\varphi_{\mathrm{Zak}}-\varphi_{\mathrm{Zak},+}(\Delta)+\varphi_{MW}) \right \}, \label{eq:popup}
\end{align}
and a similar expression for $n_{\downarrow}$. The Ramsey fringe is obtained by measuring the population in each spin component as a function of the MW phase $\varphi_{MW}$.\\

For negligible $n_+$ the phase of the Ramsey fringes directly corresponds to $\varphi_{\mathrm{Zak}}-\varphi_{\mathrm{Zak},-}(\Delta)$. For any finite value of $n_+$, we can recast Eq.~(\ref{eq:popup}) in the form $A+B\cos(\varphi_{Exp}+\varphi_{MW})$ and solve for $\varphi_{\mathrm{Zak}}-\varphi_{\mathrm{Zak},-}(\Delta)$ using the relation $\varphi_{\mathrm{Zak},+}(\Delta)=\pi-\varphi_{\mathrm{Zak},-}(\Delta)$ and the theoretical values of $n_{\pm}$, where $\varphi_{Exp}$ is the measured phase difference between the Ramsey fringes. The resulting values of $\varphi_{\mathrm{Zak}}-\varphi_{\mathrm{Zak},-}(\Delta)$ are displayed in Fig.\,4 of the main text, and the values of $\varphi_{Exp}$ are shown in Fig.~\ref{Fig_7}.

\end{document}